\documentclass[aip,reprint]{revtex4-1}
\usepackage{amsmath}
\usepackage{SIunits,graphicx}

\input{tcilatex}
\begin{document}

\title{Polarization dependent spontaneous-emission rate of single quantum dots in photonic crystal membranes}
\author{Q. Wang}
\email{qinw@fotonik.dtu.dk}
\author{S. Stobbe}
\author{H. Thyrrestup}
\affiliation{DTU Fotonik, Department of Photonics Engineering, Technical University of Denmark, {\O }rsteds Plads 343, DK-2800 Kgs.~Lyngby, Denmark.}

\author{H. Hofmann}
\author{M. Kamp}
\author{T. Schlereth}
\author{S. H\"ofling}
\affiliation{Technische Physik, Universit\"at W\"urzburg, Am Hubland, D-97074 W\"urzburg, Germany.}

\author{P. Lodahl}
\email{pelo@fotonik.dtu.dk}
\affiliation{DTU Fotonik, Department of Photonics Engineering, Technical University of Denmark, {\O }rsteds Plads 343, DK-2800 Kgs.~Lyngby, Denmark.}

\date{\today}

\begin{abstract}
We have measured the variation of the spontaneous emission rate with
polarization for self-assembled single quantum dots in two-dimensional
photonic crystal membranes. We observe a maximum anisotropy factor of $6$ between the decay
rates of the two bright exciton states. This
large anisotropy is attributed to the substantially different projected local
density of optical states for differently oriented dipoles in the photonic
crystal.
\end{abstract}

\pacs{78.47.da, 78.55.Cr, 78.67.Hc}
\keywords{quantum dot, photonic crystal, time-resolved measurement,
spontaneous emission}
\maketitle




In the past few decades, there has been considerable interest in applying
photonic crystals (PCs) for controlling the spontaneous emission (SE) of embedded
emitters, which may find applications in diverse areas such as quantum information science, efficient lasers and LEDs, and for energy harvesting. Originally proposed by Yablonovitch in 1987\cite{yabl}, the experimental progress has been delayed due to the lack of sufficiently high quality emitters and PCs. The first experimental demonstrations of spontaneous emission control have appeared within the last five years using colloidal quantum dots or dye molecules in 3D opal PCs \cite{peter1,niko1,niko2} and self-assembled quantum dots (QDs) or quantum wells in 2D photonic crystal membranes (PCMs) \cite{fuji,finl1,finl2,jeppe}. The latter technology has proven very successful due to the excellent optical properties of self-assembled QDs \cite{jeppe2}, the ability to optically address single QDs \cite{finl2}, and the strongly modified optical local density of states (LDOS) in PCMs \cite{koen1}. Recently it was theoretically proposed that the spontaneous emission (SE) rate in a PC can be highly anisotropic depending on the orientation of the transition dipole moment of the emitter \cite{koen3}, which may be employed to enhance effects of quantum interference between the two radiating states \cite{agar} of relevance for quantum information applications. Here we experimentally demonstrate such a pronounced anisotropy by carrying out time- and polarization-resolved spontaneous emission measurements on a single QD addressing two orthogonally polarized bright exciton states. In this process, we probe the anisotropy of the vacuum electromagnetic field in the PCMs, which was not addressed experimentally previously.

When optically exciting a QD, choosing the sample growth direction $
[001]$ as the quantization axis (z) for angular momentum, one lifts an
electron ($S_{e,z}=\pm \frac{1}{2}$) to the conduction band leaving a heavy
hole ($J_{h,z}=\pm \frac{3}{2}$) in the valence band,
which can form four possible exciton states ($\left \vert h,e\right \rangle $): $\left
\vert \frac{3}{2},-\frac{1}{2}\right \rangle $, $\left \vert -\frac{3}{2},
\frac{1}{2}\right \rangle $, $\left \vert \frac{3}{2},\frac{1}{2}\right
\rangle $,$\left \vert -\frac{3}{2},-\frac{1}{2}\right \rangle $. We note
that the light holes ($J_{h,z}=\pm \frac{1}{2}$) can be neglected as the
degeneracy of the light and heavy holes is
lifted by the strain causing the QDs \cite{baye}. The four
exciton states are categorized into two groups according to the
values of their total angular momentum: bright states ($J_{z}=\pm 1$) and
dark states ($J_{z}=\pm 2$), where only the bright states are
optically active. Due to the reduced symmetry of self-assembled QDs and
anisotropic exchange interactions, the two bright states are separated in
energy (typically 0-30 $\micro$eV) \cite{stev} and usually denoted as X or Y
states according to their dipole orientations ($[110]$ or [$1\overset{-}{1}0$]). The QD spontaneous emission decay curves are in general bi-exponential, where the fast component, which is considered here, is due to recombination of the bright exciton transitions while the slow component is due to dark state recombination mediated by spin-flip processes \cite{jeppe3}. Polarization resolved spontaneous-emission measurements enable addressing each of the orthogonally polarized bright exciton states individually and thereby to probe the anisotropy of the vacuum
electromagnetic field in the PCM. We quantify the
polarization dependence by defining the anisotropy factor $\eta ^{\gamma
}\equiv \frac{\gamma _{X}}{\gamma _{Y}}$, where $\gamma _{X}$ ($\gamma _{Y}$%
) represents the decay rate of the X (Y) states.

\begin{figure}[ptb]
\begin{center}
\includegraphics[scale=0.7]{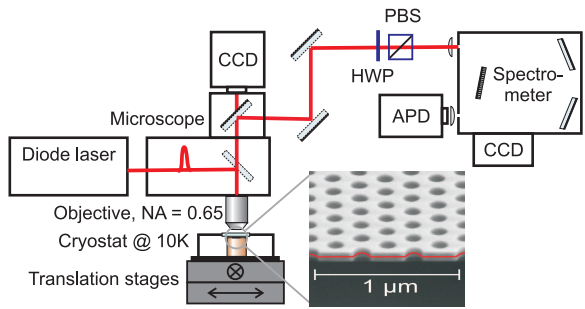}
\end{center}
\caption{(Color online) The schematic of the experimental setup. CCD: charge
coupled device camera; APD: avalanche photodiode detector; HWP: half-wave plate; PBS: polarization beam-splitter. The inset shows a scanning
electron micrograph of the sample, in which a layer of self-assembled InAs
QDs (red color) is embedded in the center of a GaAs PCM.}
\label{Fig1}
\end{figure}

The schematic of our experimental setup is illustrated in Fig.~1. The sample is a GaAs PCM with a layer of self-assembled InAs QDs of density $250$ $\micro$m$^{-2}$ embedded in the center of the membrane, see Fig.~1. It is mounted in a closed-cycle cryostat at a
temperature of $10$ K and excited from the top by a pulsed diode laser
at $780$ nm ($1.590$ eV, which is above the bandgap of GaAs)
and a repetition rate of $20$ MHz. The photoluminescence (PL) is collected
through a lens (NA = $0.65$), sent to a monochromator, and arrives either at a CCD camera for recording emission spectra or a silicon avalanche
photodiode for the time-resolved measurements. In order to facilitate polarization
resolved measurements, a polarizer consisting of a half-wave plate and a polarization beam-splitter is placed before the monochromator. The excitation
intensity used in the measurements is about $300$ mW/cm$^{2}$, which is below the exciton saturation level so that only photon emission from the ground state is observed.
The ground state emission wavelength is centered at $950$ nm ($1.305$ eV)
with an inhomogeneous broadening of $70$ meV. The resolution of
the monochromator is about $120$ $\micro$eV, which is larger than the energy
splitting between the two bright states. However, they can still be separated by their different polarization.

During our experiments, we investigated about $30$ different QDs positioned in 7 different PCMs, with the lattice parameters
ranging from $260$ nm to $320$ nm. For the sake of exploiting a pronounced
2D PC bandgap effect, we chose QDs in PCMs with $r/a=0.30$, where $r$ is
the radius of the air holes and $a$ is the lattice constant. For
comparison, we also measured decay curves of $4$ QDs positioned outside the PCMs. For each QD, the
PL was projected onto different polarization directions by changing the
orientation of the half-wave plate before the monochromator.

\begin{figure}[ptb]
\begin{center}
\includegraphics[scale=0.7]{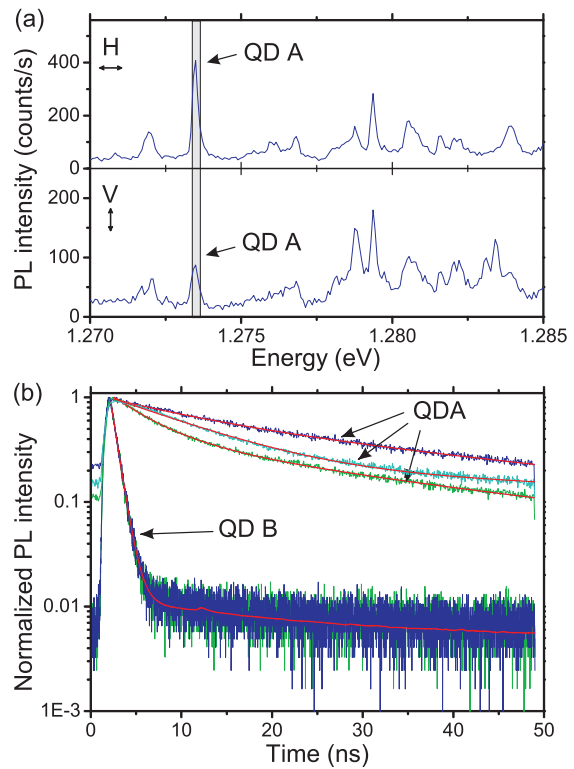}
\end{center}
\caption{(Color online) (a). PL spectra for QDs positioned in a PCM ($a=320$ nm)
measured at H or V polarizations displaying single QD lines. The shaded area represents the
resolution of the spectrometer. (b). Three decay curves for QD A (inside
PCM, emission energy $1.274$ eV) corresponding to $0^{\circ }$ (blue, upper curve), $70^{\circ }$ (cyan, middle curve) or $90^{\circ }$ (green, lower curve)
polarization. Also shown are two decay curves for QD B (outside PCM, emission energy $1.267$ eV) for $
0^{\circ }$ (blue curve) and $90^{\circ }$ (green curve) polarizations that are almost on top of each other. The red lines are bi-exponential fits to the decay curves.}
\label{Fig2ab}
\end{figure}

Fig.~2(a) shows the PL spectrum of single QDs inside a PCM by recording either
horizontal ($H$) or vertical ($V$) polarizations. The spectrum is composed of
sharp emission lines originating from single QDs with linewidths limited by the resolution of the
spectrometer, as indicated by the shaded area in Fig.~2(a). Fig.~2(b) displays
typical decay curves for two QDs, where QD A is inside a PCM, and QD B is in the unpatterned substrate while being close in emission energy to QD A. Three
decay curves for QD A are displayed corresponding to different polarization components $0^{\circ }$ (i.e.~H), $70^{\circ }$, and $%
90^{\circ }$ (i.e.~V).  We clearly observe that the SE rate is strongly dependent on polarization illustrating that X and Y bright excitons decay significantly different in the PCM due to the anisotropic vacuum fluctuations experienced by the QD. For comparison, no such anisotropy is observed in the reference measurements on QD B. The SE rate is furthermore found to be strongly inhibited in the PCM with the inhibition
factors differing for X and Y. By comparing QD A and B we derive an inhibition factor of $15.8$
for the X state and $6.5$ for the Y state.

The PL intensity and decay rate obtained when probing different polarizations for QD A are
presented in Fig.~3. Polarizations H and V correspond to probing the two orthogonally
polarized bright states X and Y, while intermediate directions probe a combination of the two bright states. Note that this implies that only in the former case are the decay curves strictly bi-exponential functions. However this model turns out to model the decay curves rather well also for intermediate polarization settings, and the goodness-of-fit ($\chi ^{2}$) varying between $1.0$ and $1.4$ is found for the complete data set. The PL intensity shows a maximum
(minimum) value at H (V) polarization, which is opposite to the decay rate.
This is expected since a strong suppression of the decay rate in the plane of the PCM
results in a high emission vertically out of the membrane due to energy
redistribution \cite{fuji}. The PL intensity variation with polarization $\theta$ is observed to
follow the simple relation $I=\frac{I_X+I_Y}{2}+\frac{I_X-I_Y}{2}\cos (2\theta )$, where $I_X$ and $I_Y$ are the intensities of the $X$ and $Y$ exciton states, respectively, see Fig.~3. This can be easily understood as the result of applying polarization projection measurements on two orthogonal states.

\begin{figure}[ptb]
\begin{center}
\includegraphics[scale=0.7]{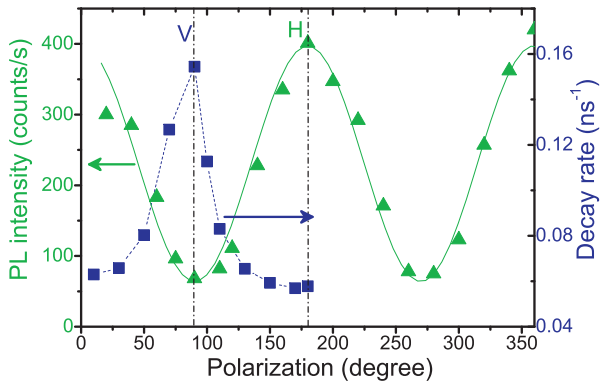}
\end{center}
\caption{(Color online) PL intensities and decay rates versus polarization
for QD A. The triangular points (square points) are experimental results
for intensities (decay rates). The solid line is the fitted result with a
cosine function, and the dashed line is a guide to the eye.}
\label{Fig3}
\end{figure}

\begin{figure}[ptb]
\begin{center}
\includegraphics[scale=0.7]{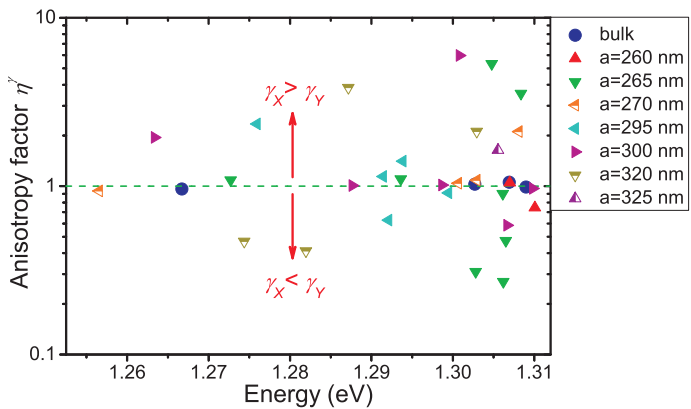}
\end{center}
\caption{(Color online) Measured anisotropy factor of decay rates between X
and Y states. The triangular points represent QDs inside
PCMs (with 7 different lattice parameters), the circular points represent QDs outside PCMs, and the dashed
horizontal line separates regions $\protect \gamma _{X}>\protect \gamma _{Y}$
and $\protect \gamma _{X}<\protect \gamma _{Y}$.}
\label{Fig4}
\end{figure}

Fig.~4 shows the anisotropy factor of decay rates for all the measured QDs measured on PCMs with various values of the lattice constant. Note that in  all measurements presented in the present manuscript the QD emission was within the 2D photonic bandgap of the PCMs \cite{koen1}.
Large variations are observed between the individual QDs in the PCM with a maximum value of about $6$. This directly demonstrates the large anisotropy of the vacuum electromagnetic field in a PC that was theoretically proposed in Ref.~\cite{koen3}. This anisotropy gives rise to substantial differences in the projected LDOS leading to the different decay dynamics of X and Y exciton states. For comparison, reference QDs in a bulk substrate showed no anisotropy in the decay rates for the two orthogonally polarized states.

To conclude, we have systematically measured the polarization dependent SE
rate for self-assembled single QDs inside PCMs and obtained a maximum
anisotropy factor of decay rate between the X and Y states of $6$. Our
measurement results demonstrate the large anisotropy of the vacuum electromagnetic field
inside PCMs \cite{koen1,koen3}, which is a crucial condition for achieving
quantum interference between two closely lying energy levels \cite{agar} that could enable demonstration of fascinating phenomena, such as lasing
without inversion \cite{imam} or
quantum beats \cite{koch}. Therefore, our experiment is not
only vital in realizing complete control of the SE of single QDs with PCs,
but also enables fundamental quantum optics experiments with practical systems.

\begin{acknowledgments}
We thank T. Lund-Hansen and M. L. Andersen for help during the experiment,
and we gratefully acknowledge financial supports from the Danish Research
Council (FTP grant 274-07-0459 and FTP/FNU grant 272-09-0159).
\end{acknowledgments}

\end{document}